\documentclass[twocolumn]{aastex631}

\newcommand{\msun}{\ensuremath{M_{\odot}}}

\graphicspath{{./}}

\begin{document}

\title{Prospects for a precise equation of state measurement from Advanced LIGO and Cosmic Explorer}

\correspondingauthor{Daniel Finstad}
\email{dfinstad@berkeley.edu}

\author[0000-0002-8133-3557]{Daniel Finstad}
\affiliation{Institute for Nuclear Theory, University of Washington, Seattle, WA 98195-1550, USA}
\affiliation{University of California, Berkeley, CA 94720, USA}
\affiliation{Lawrence Berkeley National Laboratory, Berkeley, CA 94720, USA}

\author[0000-0001-6699-1300]{Laurel V. White}
\author[0000-0002-9180-5765]{Duncan A. Brown}
\affiliation{Department of Physics, Syracuse University, Syracuse NY 13244, USA}

\date{\today}

\begin{abstract}

Gravitational-wave observations of neutron star mergers can probe the nuclear equation of state by measuring the imprint of the neutron star's tidal deformability on the signal.
We investigate the ability of future gravitational-wave observations to produce a precise measurement of the equation of state from binary neutron star inspirals.
Since measurability of the tidal effect depends on the equation of state, we explore several equations of state that span current observational constraints.
We generate a population of binary neutron stars as seen by a simulated Advanced LIGO--Virgo network, as well as by a planned Cosmic Explorer observatory.
We perform Bayesian inference to measure the parameters of each signal, and we combine measurements across each population to determine $R_{1.4}$, the radius of a $1.4\msun$ neutron star.
We find that with 321 signals the LIGO--Virgo network is able to measure $R_{1.4}$ to better than 2\% precision for all equations of state we consider, however we find that achieving this precision could take decades of observation, depending on the equation of state and the merger rate.
On the other hand we find that with one year of observation, Cosmic Explorer will measure $R_{1.4}$ to better than 0.6\% precision.
In both cases we find that systematic biases, such as from an incorrect mass prior, can significantly impact measurement accuracy and efforts will be required to mitigate these effects.

\end{abstract}

\section{Introduction}

The observation of the binary neutron star merger GW170817 by the Advanced LIGO--Virgo network provided the first constraints on the dense matter equation of state through gravitational waves~\citep{LIGOScientific:2017vwq,LIGOScientific:2018hze}. Prior to this event, constraints on the equation of state were fairly broad, with observations and experiments allowing a wide range of neutron star radii (see ~\cite{Lattimer:2012nd} and references therein for a review). Now with new constraints provided by the many studies of GW170817 and its electromagnetic counterpart (see e.g. \cite{LIGOScientific:2018cki,Breschi:2021tbm,Capano:2019eae,Coughlin:2018miv,De:2018uhw,Dietrich:2017aum,Pang:2022rzc,Radice:2018ozg}), as well as recent observations of pulsars and measurements of neutron skin thickness~\citep{Miller:2021qha,Raaijmakers:2019qny,Reed:2021nqk,Riley:2019yda,Riley:2021pdl}, the plausible range of equations of state has narrowed considerably, with the consensus tending toward those that predict a radius $R_{1.4}$ for a canonical $1.4\msun$ neutron star of between 10.5 km and 13 km.

As the constraint window continues to narrow, future measurements of the equation of state will require very high ($\sim2\%$) precision in order to distinguish between plausible models, and it remains to be seen whether such an extraordinary precision will be achievable by the current generation of gravitational-wave detectors, or if the upcoming third-generation detectors will be necessary. The next observing runs of the Advanced LIGO network will operate with increasing sensitivity, approaching design sensitivity for the fifth observing run (scheduled to begin in 2025) and beyond~\citep{Aasi:2013wya}. Third-generation detectors, such as Cosmic Explorer in the US and Einstein Telescope in the EU, are expected to achieve another order of magnitude improvement in sensitivity as compared to Advanced LIGO. Cosmic Explorer is expected to come online in the mid-2030s, at which point it will be sensitive to the entire population of merging binaries out to a redshift of $z=1$~\citep{Evans:2021gyd}. The ability of either detector network to reach the desired precision will depend on the merger rate of binary neutron star systems in the local universe, as well as on the equation of state itself.

The ability to measure the equation of state in gravitational-wave signals depends very sensitively on the equation of state itself, because the range of plausible models predict varying amounts of information in an inspiral waveform. Gravitational-wave signals from coalescing neutron stars carry information about the equation of state as a result of the tidal deformation that the stars' gravitational fields produce in one another. Specifically, the quadrupole moment $Q_{ij}$ of one neutron star is related to the tidal field $\mathcal{E}_{ij}$ of the other neutron star according to $Q_{ij} = -\lambda\mathcal{E}_{ij}$, where $\lambda$ is the tidal deformability of the neutron star~\citep{Flanagan:2007ix}. The tidal deformability is dependent on the equation of state and is commonly expressed in dimensionless form as
\begin{equation}
    \Lambda=\frac{2}{3}k_{2}\left(\frac{Rc^{2}}{Gm}\right)^5
\end{equation}
where $k_{2}$ is the tidal Love number. $R$ and $m$ are the radius and mass of the neutron star, respectively. The energy expended in deforming the stars results in a different phase evolution of the gravitational waveform as compared to a signal with non-deforming bodies. An equation of state that has a large $\Lambda$ for a given mass is said to be ``stiff", and will generally correspond to a larger radius as the neutron star is more able to hold itself up against its own gravity. A stiff equation of state produces a larger effect on the gravitational-wave phasing and is therefore more measurable. Conversely, a ``soft" equation of state will have a smaller $\Lambda$ and radius for a given mass, and produces a less measurable effect in a gravitational-wave signal.

Given the inherent difficulty of measuring the equation of state through gravitational waves, an established method of improving constraints is to combine multiple observations of a single universal quantity of interest, such as $R_{1.4}$. Previous works have used various implementations of this method to estimate the measurability of the equation of state for a given signal population or detector network:
\begin{itemize}
    \item \cite{Lackey:2014fwa} combine 20 signals in a simulated LIGO--Virgo network at design sensitivity for several choices of equation of state and constrain the neutron star radius to within 1 km.
    
    \item \cite{Agathos:2015uaa} combine 200 signals in a LIGO--Virgo network and find that a catalog of $>100$ signals is sufficient to distinguish tidal deformability measurements between soft, moderate, and stiff equations of state. The authors also find an incorrect mass prior induces a systematic bias in the equation of state measurement.

    \item \cite{Vivanco:2019qnt} combine 20 signals in a LIGO--Virgo network and project that the 8 loudest signals will constrain $R_{1.4}$ to within 10\%.

    \item \cite{Pacilio:2021jmq} combine 20 signals in a LIGO--Virgo network and an Einstein Telescope observatory. They find that Einstein Telescope is able to distinguish between tidal deformabilities for similarly soft equations of state, whereas LIGO--Virgo cannot.

    \item \cite{Chatziioannou:2021tdi} combine signals in different detectors according to their sensitivity and an estimated merger rate for binary neutron stars, finding their most precise radius constraint of $0.05-0.2$ km with 4 years
    of observation in another planned next-generation detector, Voyager.
    % of observation in a third-generation Voyager detector.
\end{itemize}

In this paper we build upon these previous works to produce a forecast for a precise measurement of the equation of state using both a LIGO--Virgo network operating at design sensitivity, as well as a simulated Cosmic Explorer. We make our forecast using an astrophysically motivated population of simulated binary neutron star signals for each detector network. We investigate measurability for three choices of equation of state that span the range of the most recent constraints from gravitational-wave and electromagnetic observations, as well as nuclear experiments. We perform full Bayesian parameter estimation on each signal to recover their source parameters, and we combine measurements of $R_{1.4}$ across each population to produce a precise constraint. We use a collection of equations of state built from chiral effective field theory as a prior in our parameter estimation analysis which allows us to transform each measurement to $R_{1.4}$ without incurring the additional uncertainty that comes from other methods (e.g. quasi-universal relations). Finally, we investigate the impact of an incorrect choice of mass prior in the context of a precise equation of state measurement.

The rest of this paper is organized as follows: In Section~\ref{sec:tidal_meas} we give background information on measuring tidal deformability in gravitational-wave inspiral signals and describe the equations of state used in our analysis. Section~\ref{sec:population} gives details of the simulated binary neutron star signals we analyze, and Section~\ref{sec:pe} outlines the parameter estimation framework we use to recover the signals and combine measurements across the population. In Section~\ref{sec:results} we present our combined $R_{1.4}$ measurement for each equation of state and detector network we consider before concluding in Section~\ref{sec:conclusion}.

\section{Measurement of tidal deformability}\label{sec:tidal_meas}

Information about the nuclear equation of state is encoded in the tidal deformability of the neutron stars, which changes the orbital evolution of the binary and hence the radiated gravitational waves~\cite{Flanagan:2007ix}. In the frequency-domain, an inspiraling gravitational-wave signal can be expressed as
\begin{equation}
    h(f)=\mathcal{A}f^{-7/6}\exp[i(\psi_\mathrm{pp}+\psi_\mathrm{tidal})]
\end{equation}
where $\mathcal{A}$ is the waveform amplitude, $\psi_{\rm{pp}}$ is the point-particle contribution to the phasing (including spin effects), and $\psi_{\rm{tidal}}$ is the contribution to the phasing from tidal effects. At leading order, the tidal phasing is determined by the \emph{effective} tidal deformability $\tilde\Lambda$, defined as

\begin{equation}
    \tilde\Lambda=\frac{16}{13}\frac{(12q+1)\Lambda_{1}+(12+q)q^{4}\Lambda_{2}}{(1+q)^{5}}
\end{equation}
where $q=m_{2}/m_{1}\leq1$ is the mass ratio. Then the leading order tidal phasing in a gravitational-wave inspiral is proportional to $\tilde\Lambda$
\begin{equation}\label{eq:tidal_phase}
    \psi_{\mathrm{tidal}} \propto \tilde\Lambda f^{5/3}.
\end{equation}

By measuring the phase of a gravitational-wave signal we can determine the amount of deformation occurring in the neutron stars, giving us insight into the structure of the dense matter comprising their interior. However, as is clear from Equation~\ref{eq:tidal_phase} the tidal effect depends strongly on the gravitational-wave frequency. It has previously been found that tidal information in a signal only becomes measurable for $f\gtrsim400$ Hz~\citep{Harry:2018hke}, with the largest effect occurring just before merger at frequencies $f>1$ kHz. Measuring this higher frequency portion of a signal presents a significant challenge because of the chirping nature of the signal; a binary neutron star signal may spend minutes (or hours, in the case of Cosmic Explorer) in the sensitive band of the detectors, but only a fraction of a second at frequencies above 400 Hz. Given the extreme difficulty in extracting information from this last moment before merger, differences in tidal measurability between soft and stiff equations of state can be significant.

\begin{figure}[ht]
\includegraphics[width=0.45\textwidth]{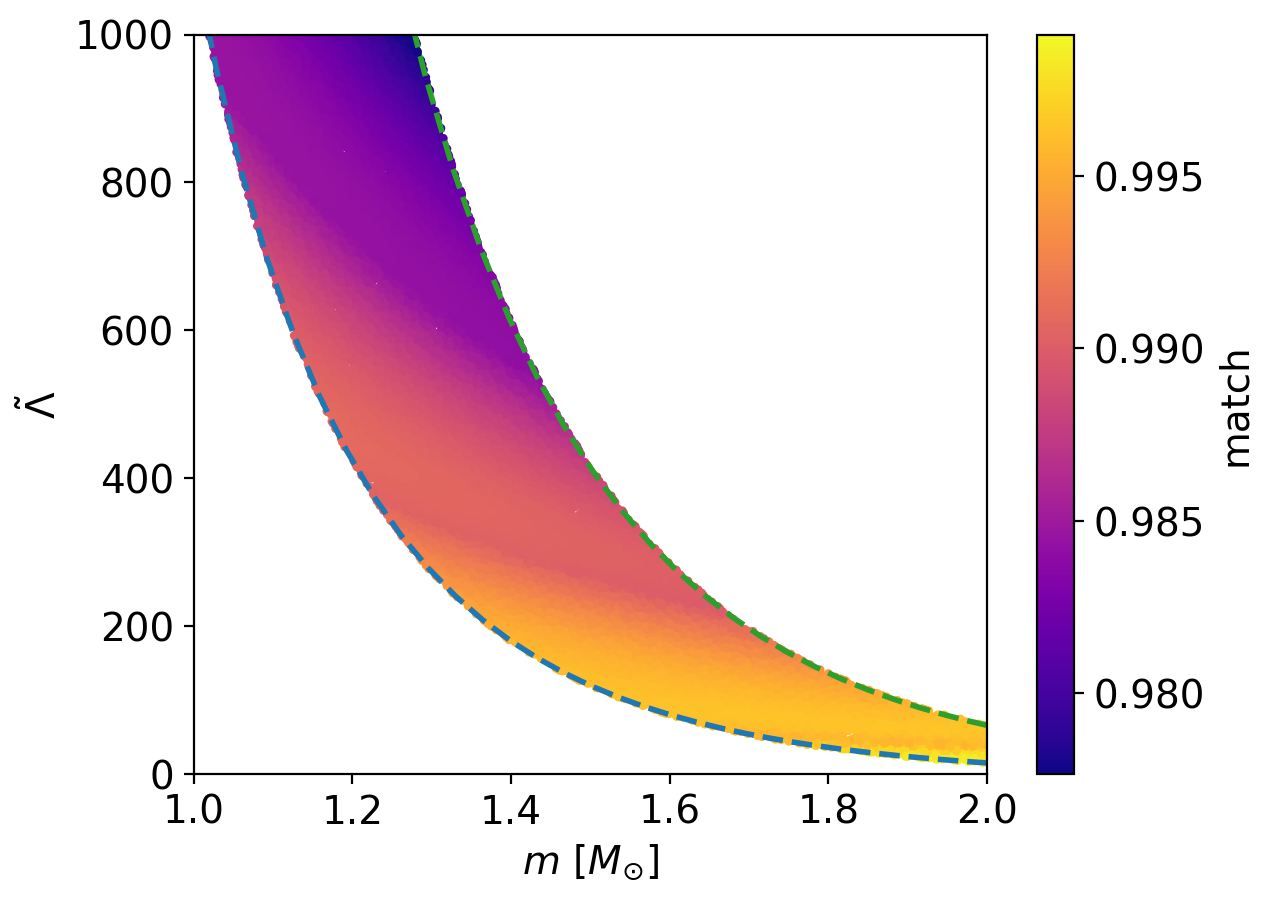}
\caption{Match between gravitational waveforms for equal mass binaries with and without tidal deformability included. The match is calculated as the noise-weighted overlap between the two waveforms in the frequency range $20-2048$ Hz using the Advanced LIGO design sensitivity noise curve. Waveforms are generated using masses ranging from $1-2$\msun, and for the waveform including tidal deformability we use values of $\tilde\Lambda=\Lambda_{1}=\Lambda_{2}$ that span the range of plausible values between the soft (blue, lower curve) and stiff (green, upper curve) equations of state selected for our analysis.}
\label{fig:tidal_match}
\end{figure}

To illustrate the effect that stiffness can have on tidal measurability, in Figure~\ref{fig:tidal_match} we show the match between gravitational waveforms with tidal information included versus corresponding waveforms without tides. The match is defined as the noise-weighted overlap between the two waveforms, so a lower match is analogous to greater measurability of the tidal information. We calculate the match for different combinations of $\tilde\Lambda$ and neutron star mass $m$ for the case of an equal-mass binary seen in an Advanced LIGO detector at design sensitivity. We plot also the functional $m-\tilde\Lambda$ relationship for two equations of state representing the approximate bounds of the plausible range. Both equations of state pass through regions of differing match owing to the mass dependence of the tidal deformability, but the stiff equation of state consistently lies in regions with substantially lower match implying a greater measurability. This is especially true for neutron star masses below about 1.6\msun~ where the plausible region spans a wider range of $\tilde\Lambda$ for a given mass.

\begin{figure}[ht]
\includegraphics[width=0.45\textwidth]{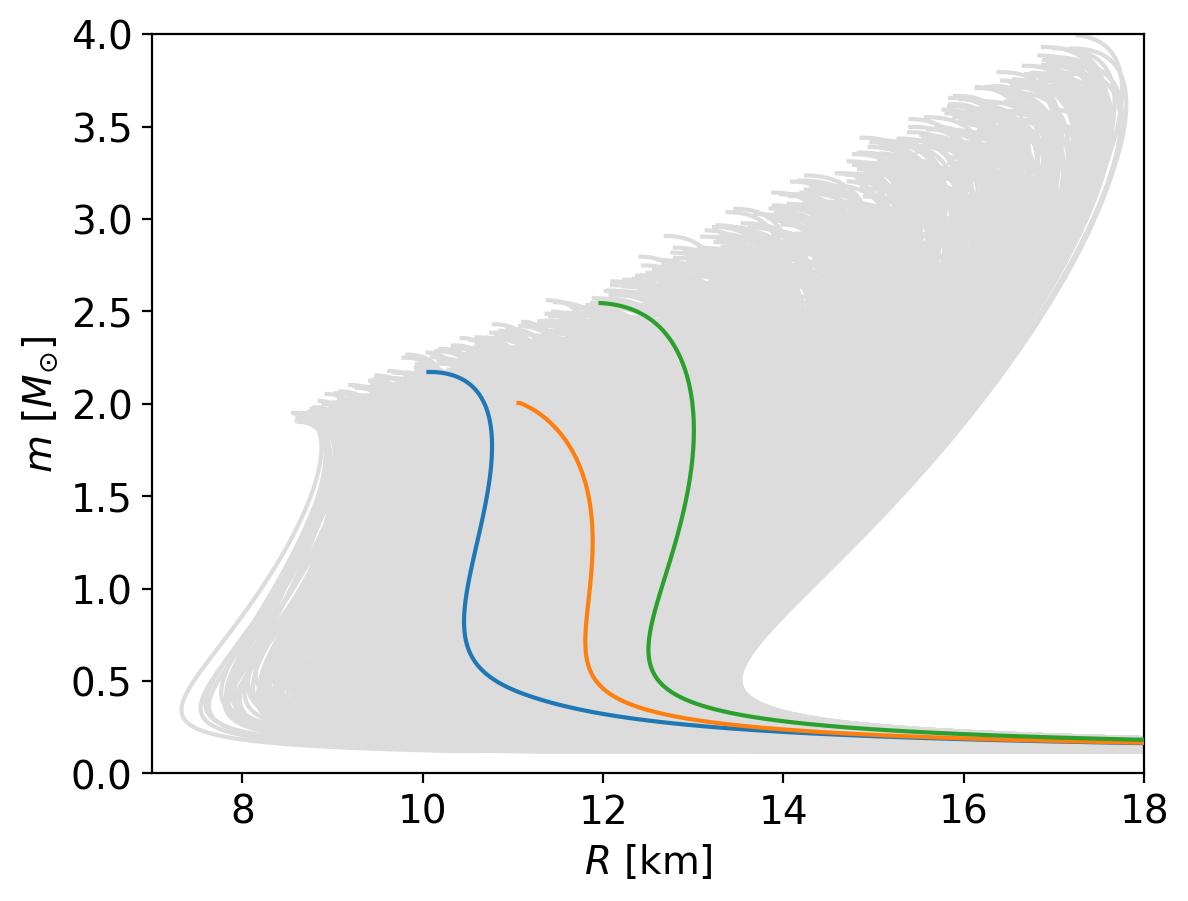}
\caption{Mass-radius curves for the soft (blue), medium (orange), and stiff (green) equations of state we consider in our analysis. These equations of state were chosen to approximately span the range of currently allowed stiffness based on observational constraints. Also shown in grey is the full set of 2000 equations of state that we use as a prior distribution in the parameter estimation analysis of our simulated signals.}
\label{fig:eosplaceholder}
\end{figure}

Since our ability to measure the equation of state depends on its stiffness, we select three equations of state for our analysis that span the range allowed by current observational constraints, hereafter referred to as ``soft," ``medium," and ``stiff". We require that each equation of state support a maximum neutron star mass of at least 2\msun, and for simplicity we do not select equations of state that include a phase transition. The equations of state we consider are selected from a set of 2000 equations of state constructed from chiral effective field theory, which uses an order-by-order inclusion of nucleon interactions governed by pion-exchange~\citep{Epelbaum:2008ga,Machleidt:2011zz}. It is known that chiral effective field theory will break down at very high densities, though studies have suggested that the theory is valid up to between $1-2$ times nuclear saturation density (see \cite{Drischler:2021kxf} and references therein). As a conservative choice, for our analysis we use equations of state calibrated up to nuclear saturation density. In Figure~\ref{fig:eosplaceholder} we show the mass-radius curves for the three equations of state we consider, as well as the full set of 2000 equations of state that we use as a prior distribution (discussed in greater detail in Section~\ref{sec:pe}).

\section{Simulated signals}\label{sec:population}

To test the ability of gravitational-wave observatories to measure the equation of state, we generate a simulated population of binary neutron star mergers, and project their gravitational waveforms onto a detector network. For each detector network (LIGO--Virgo and Cosmic Explorer) we generate three copies of our simulated population, one for each equation of state we consider. Within a given population, every neutron star is assigned a tidal deformability according to its mass and the equation of state associated with the population.

Electromagnetic observations of binary neutron star systems are fairly limited in number, but it has been found previously that the mass distribution of neutron stars in these systems is well described by a Gaussian centered near $1.4\msun$~ with a standard deviation of $\sigma_{m}=0.05\msun$~(see e.g.~\cite{Kiziltan:2010} and \cite{Ozel:2012ax}). Neutron star spins in these systems have also been observed to be small; to date the fastest known pulsar in a double neutron star system, J0737-3039A, has a spin period of 22.70 ms~\citep{Burgay:2003jj} which corresponds to a dimensionless spin of only $\chi\sim0.05$. In line with these results, we generate our population with masses drawn from a Gaussian centered at $1.4\msun$~ with $\sigma_{m}=0.05\msun$, and spins in the direction of the orbital angular momentum drawn from a zero-mean Gaussian with $\sigma_{\chi}=0.02$. Sky locations are distributed uniformly across the sky, and the inclination and orientation of the binary systems is distributed uniformly on the sphere. For signals analyzed with the LIGO--Virgo network, distances are drawn uniformly in volume in the range [20, 585] Mpc. Signals analyzed with Cosmic Explorer have distances drawn uniformly in volume in the range [20, 1100] Mpc. Both upper bounds represent the largest distance at which an optimally oriented equal-mass binary of total mass $4\msun$ would produce a single detector signal-to-noise of 8 and 100 for the LIGO--Virgo and Cosmic Explorer detectors, respectively. The signal-to-noise threshold of 100 for Cosmic Explorer was chosen to produce a population of signals that is representative of what is expected from one year of observation~\citep{Evans:2021gyd}.

For the LIGO--Virgo network, we simulate a three-detector network representing the LIGO Hanford, LIGO Livingston~\citep{TheLIGOScientific:2016agk,Buikema:2020dlj}, and Virgo~\citep{TheVirgo:2014hva} detectors. Each LIGO--Virgo detector is simulated at its design sensitivity by injecting simulated signals into Gaussian noise colored by the design power spectral density for each detector~\citep{Aasi:2013wya}. Cosmic Explorer is still in its design phase and does not have a final configuration or site location determined yet, although potential sites in the United States include locations in Utah or Idaho. For simplicity we use a hypothetical Cosmic Explorer detector at the same location and orientation as the LIGO Hanford detector. We choose the 40 kilometer arm length configuration optimized for detection of coalescing binaries for our analysis, and signals are injected into Gaussian noise colored by the corresponding design power spectral density~\citep{CE:NoiseCurves}.

LIGO--Virgo signals are pre-filtered via a network matched-filter signal-to-noise calculation~\citep{Allen:2005fk} to select the subset with signal-to-noise $\rho_{\mathrm{mf}}>13.85$, which is equivalent to a signal-to-noise of 8 in each detector. The resulting LIGO--Virgo population contains 321 signals with signal-to-noise ratios that range from about 10 to 73. Cosmic Explorer signals are filtered to require $\rho_{\mathrm{mf}}>100$, and the resulting population contains 346 signals with signal-to-noise ranging from 97 to 790.

All simulated signals are generated using the \texttt{IMRPhenomD\_NRTidal} waveform approximant~\citep{Husa:2015iqa,Khan:2015jqa,Dietrich:2017aum}, which is a frequency domain waveform available through the LIGO Algorithm Library~\citep{2020ascl.soft12021L}. The waveform uses a phenomenological model tuned to numerical relativity data to capture the inspiral, merger, and ringdown portions of a binary coalescence.

\section{Parameter estimation}\label{sec:pe}

In general, under the assumption of Gaussian noise characterized by a power spectral density $S(f)$, the likelihood of obtaining detector data $d$ given the presence of a gravitational waveform $h(\theta)$ is
\begin{equation}
    \mathcal{L}(d|\theta)\propto\exp\left[-\frac{1}{2}\left<d-h(\theta)|d-h(\theta)\right>\right],
\end{equation}
where
\begin{equation}
    \left<a|b\right>=4\mathfrak{R}\int_{f_{\mathrm{min}}}^{f_{\mathrm{max}}}\frac{\tilde{a}^{*}(f)\tilde{b}(f)}{S(f)}df
\end{equation}
is the noise-weighted inner product~\citep{Finn:1992xs,Chernoff:1993th} and $\theta= \left\{ \theta_{1},\theta_{2},\ldots,\theta_{n} \right\}$ is the set of intrinsic and extrinsic parameters defining the waveform as seen by the detector. In evaluating this likelihood, we can obtain estimates of the gravitational-wave parameters $\theta$ through the joint posterior probability distribution
\begin{equation}
    p(\theta|d)\propto\mathcal{L}(d|\theta)p(\theta),
\end{equation}
where $p(\theta)$ is the assumed prior probability distribution of the parameters. Then the marginal posterior probability distribution for an individual parameter is obtained by integrating the joint posterior over all nuisance parameters. For instance, the marginalized posterior distribution for $\theta_{1}$ is
\begin{equation}
    p(\theta_{1}|d)=\int p(\theta|d)~\mathrm{d}\theta_{2}\mathrm{d}\theta_{3}\ldots\mathrm{d}\theta_{n}.
\end{equation}

We use \textit{PyCBC Inference}~\citep{Biwer_2019} with the parallel-tempered version of the \texttt{emcee} sampler~\citep{Foreman_Mackey_2013,Vousden_2015,2010CAMCS...5...65G} to sample the parameter space and produce marginalized posterior distributions for the source parameters. To help speed convergence we employ the relative likelihood model available in \textit{PyCBC Inference} which uses an approximation to the full resolution likelihood near its peak, and has been shown to produce comparable parameter estimates to non-relative models~\citep{Cornish:2010kf,Zackay:2018qdy,Finstad:2020sok}. For signals analyzed in the LIGO--Virgo network we include frequencies above a low-frequency cutoff of 20 Hz, and for Cosmic Explorer signals we use frequencies above 7 Hz. All signals are analyzed up to a high-frequency cutoff of 2048 Hz.
We sample in source-frame component masses, component spins along the direction of the orbital angular momentum, sky location, distance, geocentric time of coalescence, inclination, polarization angle, and equation of state. For each of these parameters we use a prior distribution that matches the corresponding population distribution (as described in Section~\ref{sec:population}) with the exception of the equation of state, where our prior distribution is made of a collection of 2000 equations of state designed to be roughly uniform in $R_{1.4}$ over the interval $[9, 15]$ km. Each equation of state provides a mapping between mass, radius, and tidal deformability for a neutron star. At each iteration in the analysis, a single equation of state is drawn and used to determine the tidal deformabilities of both neutron stars in the binary based on their source-frame masses. In generating a template waveform for the likelihood, source-frame masses are first converted to the detector frame through scaling by a factor of $(1+z)$, where $z$ is the cosmological redshift at the sampled distance assuming a flat $\Lambda$CDM cosmology. All template waveforms are generated using the \texttt{IMRPhenomD\_NRTidal} waveform model in order to match the simulated signals and avoid any systematic errors arising from differences between waveform models.

To produce a combined equation of state measurement across a population of signals, for each signal we transform the measured posterior distribution of equations of state into a posterior distribution of $R_{1.4}$ predicted by those equations of state. The posterior distributions of $R_{1.4}$ are then independent observations of the same universal quantity, and so they can be combined straightforwardly~\citep{DelPozzo:2013ala,Agathos:2015uaa}. For the general case of a population of $N$ signals $s_{1},s_{2},\ldots,s_{N}$, the combined $R_{1.4}$ posterior is given by 
\begin{equation}
    p(R_{1.4}|s_{1},s_{2},\ldots,s_{N})=p(R_{1.4})^{1-N}\prod_{i=1}^{N} p(R_{1.4}|s_{i})
\end{equation}
where we have used the fact that in our analysis the prior $p(R_{1.4})$ is the same for all signals.

\section{Results}\label{sec:results}

To simulate a realistic scenario of cumulatively combining each new gravitational-wave event as it occurs, we randomize the order of our signals and then combine $R_{1.4}$ posteriors one at a time to track the radius constraint (as measured by the 90\% credible interval width) as a function of the number of binary neutron star signals.

\subsection{LIGO--Virgo}

\begin{figure*}[ht]
\includegraphics[width=\textwidth]{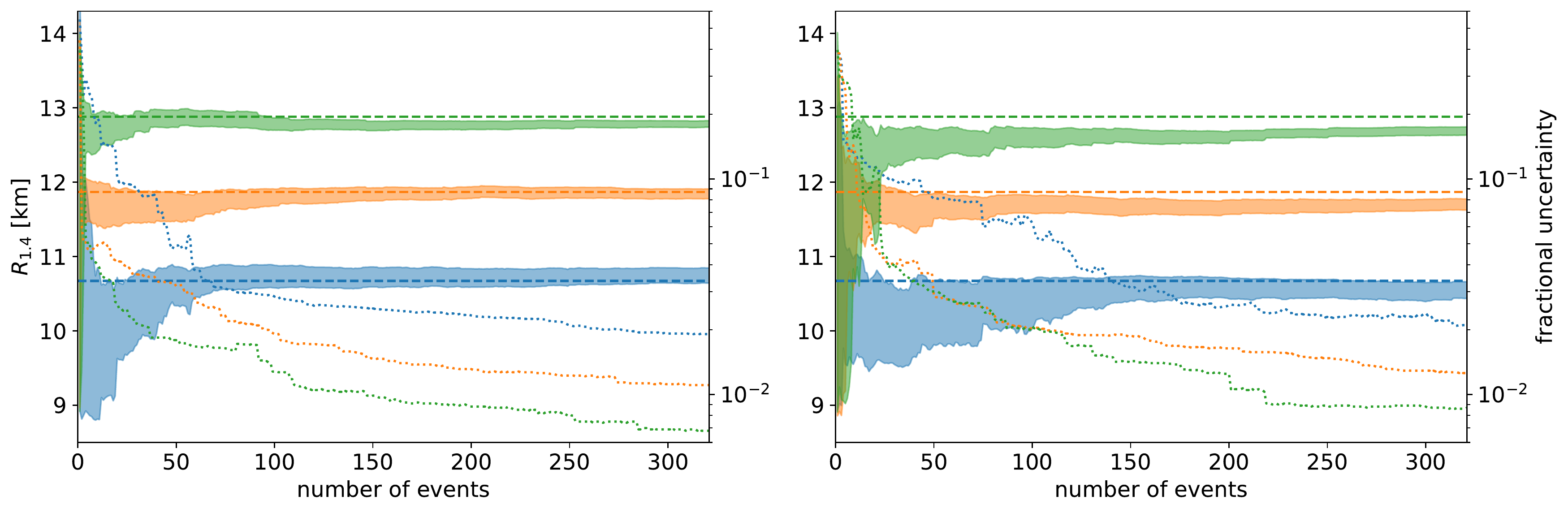}
\caption{Cumulative combined $R_{1.4}$ measurements for our simulated population in the LIGO--Virgo network. Results are shown for the soft (blue), medium (orange), and stiff (green) equations of state that we consider. Shaded regions represent the 90\% credible interval for each measurement, and the true value of $R_{1.4}$ for each equation of state is plotted as a horizontal dashed line in the appropriate color. Dotted lines show the fractional uncertainty in the measurement, calculated as the ratio of the credible interval width to the true value of $R_{1.4}$ for a given equation of state. \textit{Left:} Combined measurement when recovering signals with a Gaussian mass prior that matches the population distribution. \textit{Right:} Same as the left panel except all signals are recovered with a uniform mass prior in the interval $[1,2]\msun$. The ordering of signals is the same in both panels.}
\label{fig:hlv_2panel}
\end{figure*}

For the 321 signals in our LIGO--Virgo network analysis the combined $R_{1.4}$ constraint is shown in the left panel of Figure~\ref{fig:hlv_2panel} for each of the three equations of state we consider. We find that for each equation of state the radius constraint converges reliably toward the correct value (shown in the Figure as horizontal dashed lines). For the particular ordering shown, the constraint narrows most rapidly during the first $\sim50$ events, with the shaded uncertainty regions no longer overlapping after fewer than 10 events, although we note that our choice of ordering is arbitrary and a different choice could change the details of these features somewhat. We find that the combined constraints for the three different equations of state follow the expected hierarchy according to their stiffness; after combining all posteriors for the soft, medium, and stiff equations of state we find final 90\% credible interval widths for $R_{1.4}$ of 0.2 km, 0.13 km, and 0.09 km, respectively. As a useful gauge of measurement precision, we also convert the credible intervals to a fractional uncertainty, calculated as the ratio of the credible interval width to the true value of $R_{1.4}$ for each equation of state. The fractional uncertainties are shown in Figure~\ref{fig:hlv_2panel} as dotted lines, and it can be seen that the relationship of greater precision for stiffer equation of state is established very early in the combination process and then remains consistent through the end. After combining all posteriors the final fractional uncertainties correspond to precision of 1.9\%, 1.1\%, and 0.7\% for the soft, medium, and stiff equations of state, respectively. One could ask if our combined constraint for $R_{1.4}$ is unduly influenced by the mass distribution of our simulation population being sharply peaked around $1.4\msun$. To investigate the effect a broader mass population has on our result we repeat the analysis using a population with masses drawn uniformly in the interval $[1,2]\msun$ (and recovered with a matching uniform mass prior), but we find no significant change in the measurement precision for any of the equations of state we consider.

In previous works it has been found that imperfect knowledge of the mass distribution of the population of merging binary neutron stars can introduce a bias into a gravitational-wave measurement of the equation of state, owing to the mass dependence of the tidal deformability (see e.g.~\cite{Agathos:2015uaa,Wysocki:2020myz}). We wanted to explore the implications of this effect given the high degree of measurement precision we seek. We repeat our analysis using uniform priors in the interval $[1, 2]\msun$ for each neutron star mass. The combined $R_{1.4}$ constraint results from this analysis can be seen in the right panel of Figure~\ref{fig:hlv_2panel}, where the signal ordering is preserved from the Gaussian-prior case in the left panel. As compared to the Gaussian-prior case, we find a consistent bias toward smaller $R_{1.4}$ for all three equations of state. For the medium and stiff equations of state, the final credible interval excludes the true value of $R_{1.4}$ at high confidence, representing a disagreement at more than $3\sigma$ and $5\sigma$, respectively. We also find a slight increase in the number of events required before the uncertainty regions no longer overlap among the three equations of state, although the final 90\% credible intervals are not significantly changed.

\begin{figure}[ht]
\includegraphics[width=0.45\textwidth]{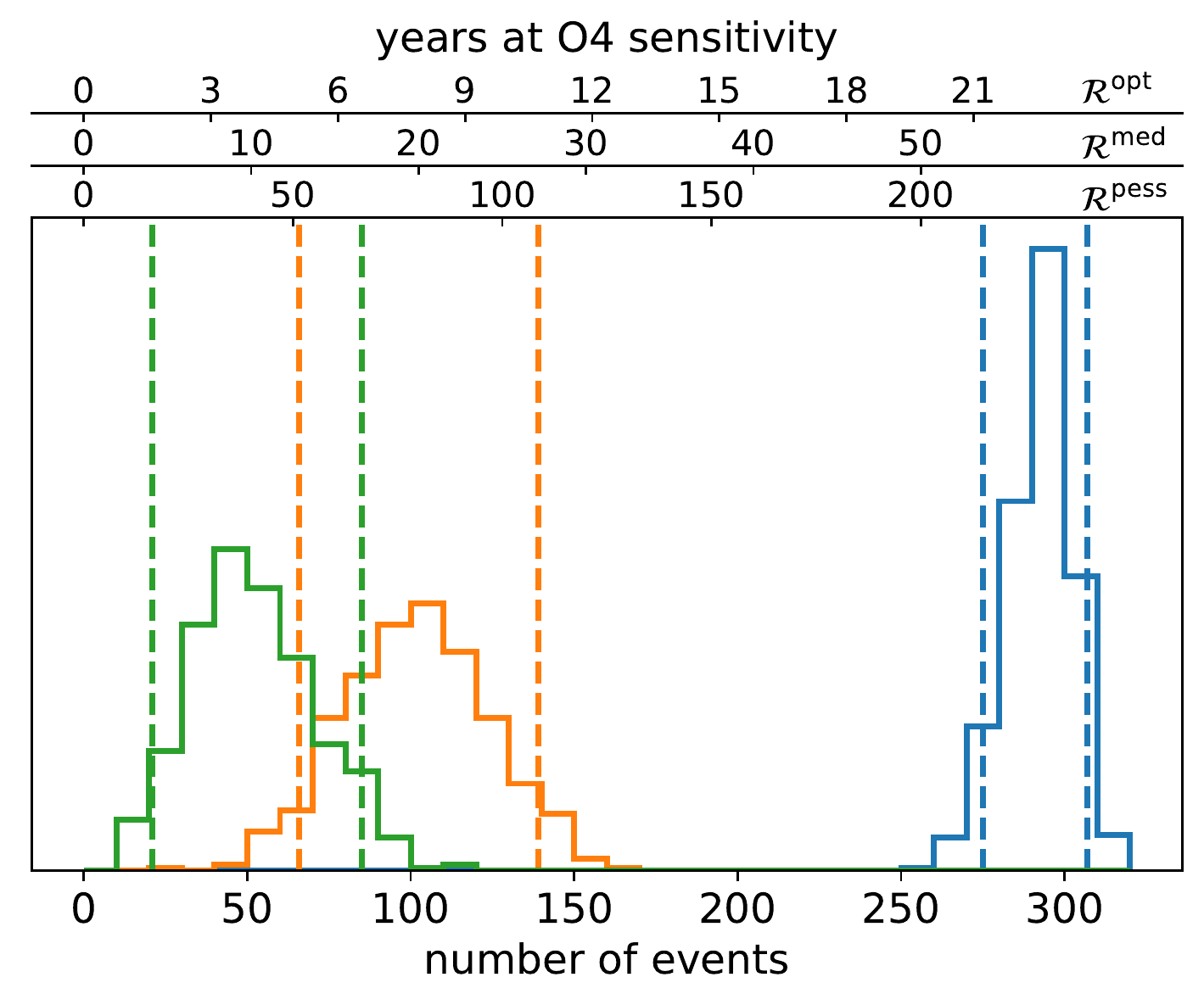}
\caption{Distribution of number of events required to reach 2\% precision for the soft (blue), medium (orange), and stiff (green) equations of state in our LIGO--Virgo analysis after 500 permutations of the combination order. Number of events are converted to years of observation at O4 sensitivity using the median estimate and upper/lower confidence limits of the binary neutron star merger rate from~\cite{Abbott:2020gyp}, shown in the top axes with labels $\mathcal{R}^{\rm pess}$ (lower limit), $\mathcal{R}^{\rm med}$ (median), and $\mathcal{R}^{\rm opt}$ (upper limit). Vertical dashed lines enclose the 90\% credible interval for each equation of state.
}
\label{fig:shuffled_dist}
\end{figure}

Our forecast shown in Figure~\ref{fig:hlv_2panel} is indeterminate for numbers of signals less than the total population, since the measurement precision from any subset of the population will depend on the specific signals included. In order to marginalize over the uncertainty in which order gravitational-wave events may occur we randomly permute the ordering of our signals 500 times, combining measurements for each ordering and determining the number of events required to reach a 2\% precision threshold. In this way we produce a probability distribution of the number of signals required to reach 2\% precision for each equation of state in our analysis, shown in Figure~\ref{fig:shuffled_dist}. We also convert signal number to years of observation at the projected sensitivity of LIGO's upcoming fourth observing run (O4) using the estimated sensitive volume $VT=0.016~\mathrm{Gpc}^{3}\mathrm{yr}$ from \cite{Aasi:2013wya}, and the median estimate as well as the upper and lower confidence limits of the binary neutron star merger rate $\mathcal{R}=320^{+490}_{-240}~\rm Gpc^{-3} \rm yr^{-1}$ from \cite{Abbott:2020gyp}. For the median merger rate, we find the LIGO--Virgo network reaches 2\% precision after $57^{+3}_{-3}$, $20^{+7}_{-7}$, and $10^{+7}_{-6}$ years (90\% confidence) at O4 sensitivity for the soft, medium, and stiff equations of state, respectively. The best-case combination we consider (high merger rate and stiff equation of state) requires only $4^{+3}_{-2}$ years at O4 sensitivity to reach 2\% precision, while the worst-case combination (low merger rate and soft equation of state) requires $228^{+12}_{-13}$ years. We note that these observation-time forecasts are upper limits, as the Advanced LIGO network is expected to undergo upgrades to its facilities which will increase its sensitivity for the fifth observing run and beyond.

\begin{figure}[ht]
\includegraphics[width=0.45\textwidth]{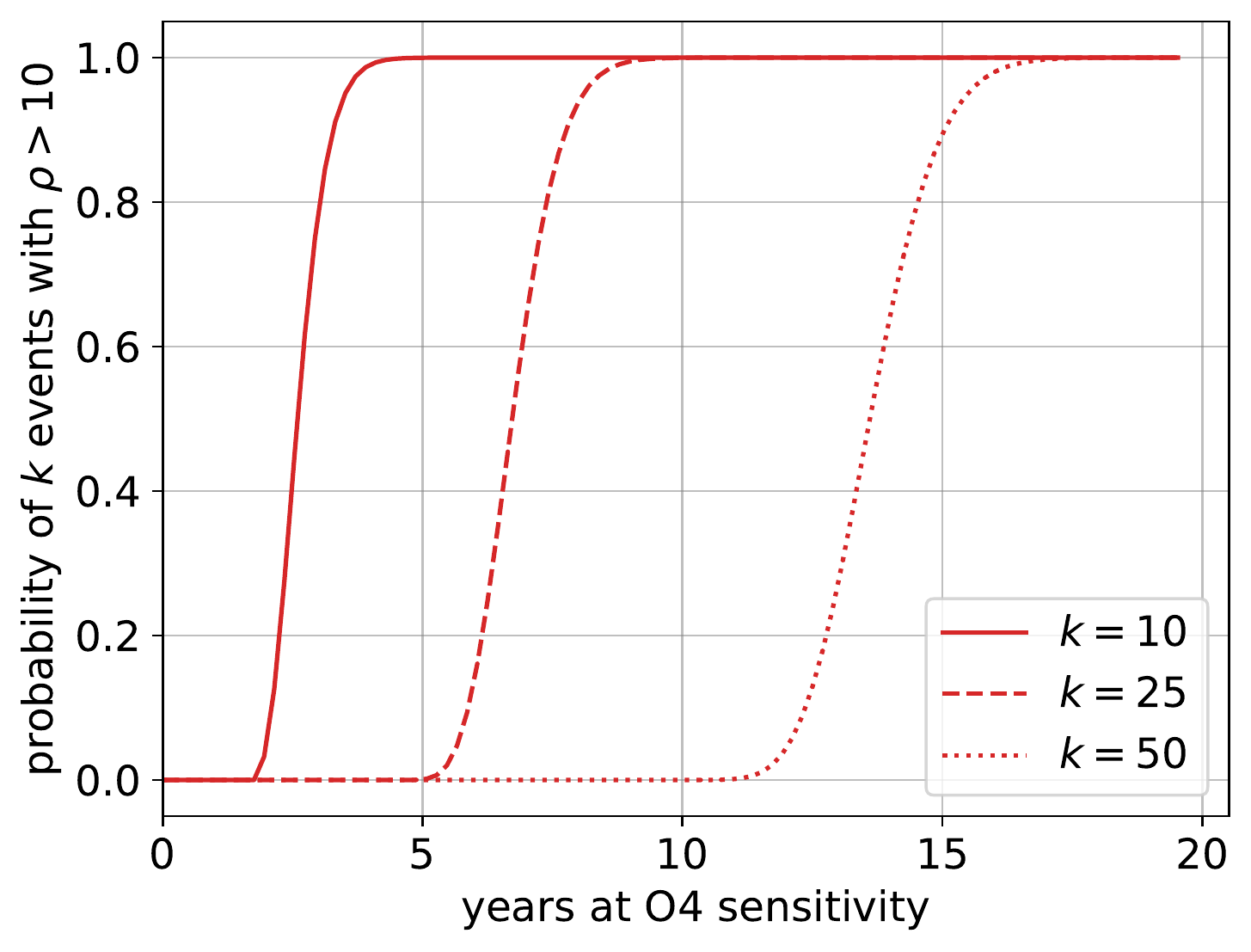}
\caption{Probability of seeing 10, 25, and 50 events with signal-to-noise $\rho>10$ over time, assuming Advanced LIGO's projected sensitivity for O4 and the median binary neutron star merger rate estimate from \cite{Abbott:2020gyp}. The probabilities for 10, 25, and 50 events are shown as solid, dashed, and dotted lines, respectively.}
\label{fig:prob_of_events}
\end{figure}

To further contextualize our LIGO--Virgo analysis results, we calculate an estimated probability of seeing different numbers of events consistent with the signals considered in this work, i.e. with signal-to-noise ratio $\rho>10$. Assuming any population of mergers in the local universe will follow the universal analytic signal-to-noise distribution described in \cite{Chen:2014yla}, we calculate this probability as a function of the total number of signals (observed or not), and then convert that to number of years at O4 sensitivity assuming the median merger rate estimate from \cite{Abbott:2020gyp} and a detection threshold signal-to-noise $\rho_{t}=9$. The calculated probabilities of seeing 10, 25, and 50 events are shown in Figure~\ref{fig:prob_of_events}. We find that under the assumptions made here, 10 signals with $\rho>10$ would almost certainly be seen in 3 years of observation, while it would require over 12 years to have any significant probability of seeing 50 such signals. We note that O4 is expected to last approximately one year, though we calculate probabilities beyond that timeline to allow for potential delays in the planned detector upgrades and to provide a lower limit for future observing runs that are expected to operate with improved sensitivity. An improved network sensitivity would effectively shift the probability curves in Figure~\ref{fig:prob_of_events} leftward by an amount equal to the factor of improvement in search volume.

\subsection{Cosmic Explorer}
\begin{figure*}[ht]
\includegraphics[width=\textwidth]{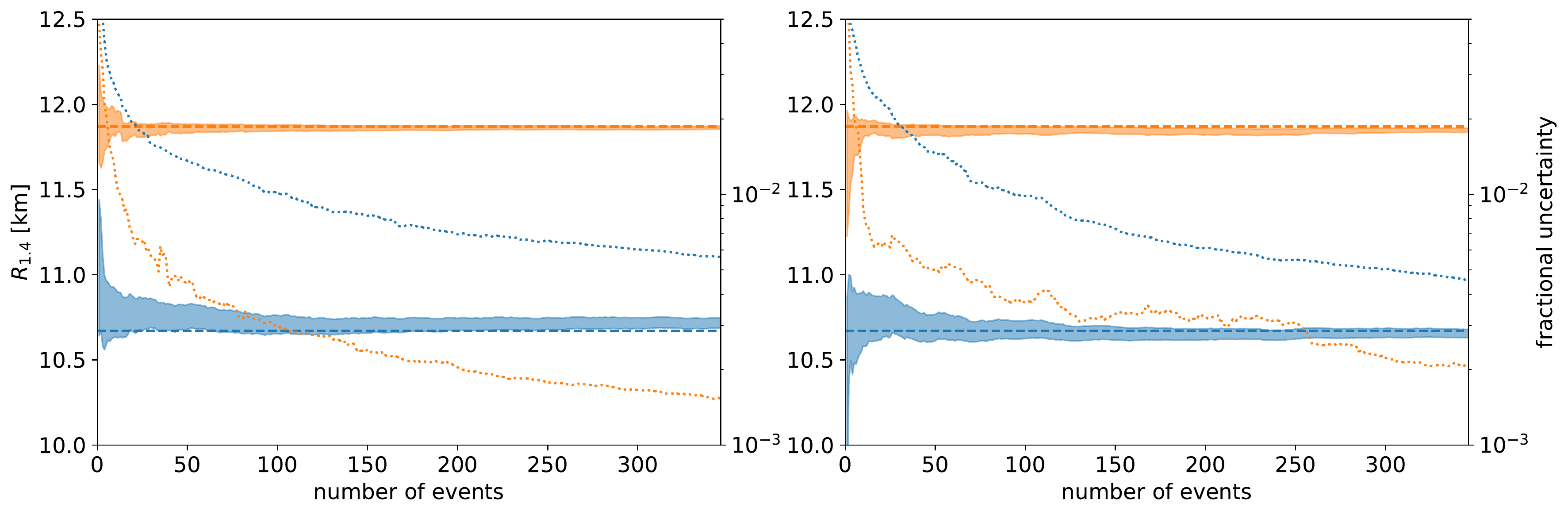}
\caption{Cumulative combined $R_{1.4}$ measurements for our Cosmic Explorer population. Results are shown for the soft (blue) and medium (orange) equations of state that we consider. Shaded regions represent the 90\% credible interval for each measurement, and the true value of $R_{1.4}$ for each equation of state is plotted as a horizontal dashed line in the appropriate color. Dotted lines show the fractional uncertainty in the measurement, calculated as the ratio of the credible interval width to the true value of $R_{1.4}$ for a given equation of state. \textit{Left:} Combined measurement when recovering signals with a Gaussian mass prior that matches the population distribution. \textit{Right:} Same as the left panel except all signals are recovered with a uniform mass prior in the interval $[1,2]\msun$. The ordering of signals is the same in both panels.}
\label{fig:ce_2panel}
\end{figure*}

Next we explore the ability of Cosmic Explorer to constrain the equation of state with our population of 346 signals that represents the expected signals with $\rho>100$ that will be seen with one year of observation. The combined $R_{1.4}$ constraint for the soft and medium equations of state are shown in the left panel of Figure~\ref{fig:ce_2panel}. We do not show a combined constraint for the stiff equation of state as we were unable to combine the full population of posteriors due to technical difficulties associated with producing such an extremely narrow distribution. As was the case in our LIGO--Virgo analysis, the combined radius constraints converge quickly toward the correct values, and the ordering of measurability according to equation of state stiffness is immediately apparent. After combining all posteriors for the soft and medium equations of state we find 90\% credible intervals of 0.06 km and 0.018 km, respectively. The corresponding fractional uncertainties for these measurements represent a precision of 0.56\% for the soft equation of state, and 0.15\% for the medium equation of state. We note that these constraint projections likely underestimate the true precision that Cosmic Explorer will achieve, as there will be an additional large population of observed signals with $\rho<100$ that will still contain measurable tidal information. While a combined equation of state measurement will almost certainly be dominated by the louder signals we consider here, it is likely that quieter signals will contribute to improve the constraint somewhat.

As with the LIGO--Virgo analysis, we investigate the effect on our Cosmic Explorer combined constraint that comes from an incorrect choice of mass prior by repeating our analysis using a uniform prior in the interval $[1,2]\msun$ for each neutron star mass. The systematic bias from an incorrect choice of prior is smaller for louder signals, so we expect that our population of Cosmic Explorer signals will suffer less from this effect. The combined $R_{1.4}$ constraint from the uniform-prior recovered signals is shown in the right panel of Figure~\ref{fig:ce_2panel}, where again the signal ordering has been preserved from the Gaussian-prior case. We find again the incorrect mass prior induces a consistent bias toward smaller $R_{1.4}$, though it is indeed a smaller effect than what we saw in the LIGO--Virgo analysis. The true $R_{1.4}$ for the soft equation of state still lies within the 90\% credible interval by the end of the combination, although the final constraint for the medium equation of state does exclude the true value. As was the case in the LIGO--Virgo analysis, we find only slight changes in the final measurement precision for both equations of state.

\subsection{Possible additional bias}
In reviewing our LIGO--Virgo and Cosmic Explorer analyses that use Gaussian mass priors that match the population distributions, we observe a potential additional bias in the combined $R_{1.4}$ measurement. In the left panels of Figure~\ref{fig:hlv_2panel} and Figure~\ref{fig:ce_2panel} it can be seen that for the soft equation of state, both the LIGO--Virgo and the Cosmic Explorer analyses overestimate $R_{1.4}$. Similarly, both analyses slightly underestimate $R_{1.4}$ for the medium equation of state. Asymmetric probabilities around the true value are generally expected with Bayesian posterior distributions, however the existence of these deflections in a consistent direction across both analyses, despite using independent populations, suggests the possibility of an additional source of systematic bias. Given that all other parameter distributions are identical between the population and the priors used in their recovery, we consider the equation of state prior to be the likeliest source of additional bias.

\section{Conclusion}\label{sec:conclusion}

We have presented an updated forecast for a precise equation of state measurement from future gravitational-wave observations of binary neutron star mergers seen by a LIGO--Virgo network and a Cosmic Explorer observatory. We performed full Bayesian inference on a simulated population of signals, and we combined individual measurements of $R_{1.4}$ into a precise combined constraint. We considered three equations of state that span the plausible range of stiffness consistent with current constraints in order to quantify measurement precision for varying degrees of tidal information.

We find that with 321 signals, the LIGO--Virgo network at design sensitivity will be able to measure $R_{1.4}$ to better than 1.9\% precision, however the observation time required to reach this precision is highly dependent on the binary neutron star merger rate and the stiffness of the equation of state. For the current median merger rate estimate from \cite{Abbott:2020gyp}, we found 2\% precision required $57^{+3}_{-3}$, $20^{+7}_{-7}$, and $10^{+7}_{-6}$ years of observation at O4 sensitivity for the soft, medium, and stiff equations of state in our analysis, respectively. We also investigated the effect of an incorrect mass prior on our results and we found that it consistently biased our measurements toward smaller $R_{1.4}$, although the overall measurement precision was mostly unaffected. In the case of the medium and stiff equation of state we considered, the bias was large enough to cause our combined constraint to exclude the true $R_{1.4}$ at high confidence. 

Our Cosmic Explorer analysis was performed on a population of 346 signals that is consistent with what is expected in one year of observations and we find that with these signals, Cosmic Explorer is able to measure $R_{1.4}$ to better than 0.56\% precision. We find that an incorrect mass prior also induces a bias toward smaller $R_{1.4}$ in our Cosmic Explorer constraint, although the effect is smaller than was seen in our LIGO--Virgo analysis due to the much louder signals in the Cosmic Explorer population. We again found that the incorrect mass prior had very little effect on the overall measurement precision with Cosmic Explorer.

One limitation of our study is that we do not account for the possibility that an identified electromagnetic counterpart could be required to confidently classify a gravitational-wave signal as a binary neutron star merger. Low-mass signals without an electromagnetic counterpart, like the probable binary neutron star signal GW190425, are open to alternative interpretations such as possibly being a neutron star--black hole system in which case their inclusion in a combined equation of state constraint that assumes both objects are neutron stars would be invalid. Analyses that produce a combined constraint including these types of signals would then rest to some extent on that assumption of their nature. Nevertheless the projection we provide here is a useful benchmark for future combined constraints.

Finally, our study highlights systematic bias as a particular concern for a precise measurement of the equation of state. Like previous works have found, we saw that imperfect knowledge of the mass distribution of neutron stars that will be seen through gravitational waves can cause a significant bias in the equation of state measurement, even with a population of relatively loud signals seen in a third-generation detector like Cosmic Explorer. However we also witnessed a possible additional bias in our analysis that assumed perfect knowledge of the mass distribution, which we attribute to a mismatch between the equation of state prior and the population distribution. While the use of a prior made up of many individual equations of state is convenient for transforming observations to a universal quantity like $R_{1.4}$, it has the downside of being unable to ever match perfectly the true population distribution which will necessarily be a single equation of state. We note that it may be possible to at least mitigate this bias through a successive narrowing of the prior used for each signal as the combined equation of state constraint gets smaller with more observations, although we leave it to a future work to investigate the application of this technique.

\begin{acknowledgements}
DF, LVW, and DAB acknowledge support from the National Science Foundation award Grant No.~PHY-2011655. DF also acknowledges support from the National Science Foundation Grant No.~PHY-2020275. DAB thanks the Kavli Institute for Theoretical Physics (KITP) for hospitality. KITP is supported in part by the National Science Foundation under Grant No.~NSF PHY-1748958. This research was supported in part through computational resources provided by Syracuse University and also made use of the Lawrencium computational cluster resource provided by the IT Division at the Lawrence Berkeley National Laboratory (Supported by the Director, Office of Science, Office of Basic Energy Sciences, of the U.S. Department of Energy under Contract No. DE-AC02-05CH11231). The authors thank Alex Nitz, Ingo Tews, and Collin Capano for helpful comments.
\end{acknowledgements}

% \bibliography{references}{}
% \bibliographystyle{aasjournal}

\end{document}